\documentclass[aps,prb,twocolumn,groupedaddress,byrevtex,showpacs,showkeys,amsfonts,amssymb,amsmath,floatfix]{revtex4-1}
\usepackage{graphicx}

\begin{document}

%Title of paper
\title{Evidence of a new low field cross-over in the vortex critical velocity of type-II superconducting thin films}

%Authors
\author{G. Grimaldi}
\email[corresponding author: ]{grimaldi@sa.infn.it}
\author{A. Leo}
\affiliation{CNR-SPIN and Dipartimento di Fisica 'E R Caianiello', Universit\`a di Salerno, Via Ponte Don Melillo, I-84084, Fisciano (SA), Italy}
\author{D. Zola}
\author{A. Nigro}
\author{S. Pace}
\affiliation{Dipartimento di Fisica 'E R Caianiello', Universit\`a di Salerno, Via Ponte Don Melillo, I-84084, Fisciano (SA), Italy}
\author{F. Laviano}
\author{E. Mezzetti}
\affiliation{Dipartimento di Fisica, Politecnico di Torino, c.so Duca degli Abruzzi 24, I-10129 Torino, Italy}

\date{\today}

%Abstract
\begin{abstract}
We measure current-voltage characteristics as function of magnetic field and temperature in Nb strips of different thickness and width. The instability voltage of the flux flow state related to the vortex critical velocity $v^{*}$ is studied and compared with the Larkin-Ovchinnikov theory. Beside the usual power-law dependence $v^{*} \approx B^{-1/2}$,  in the low field range a new cross-over field, $B_{cr1}$, is observed below which $v^{*}$ decreases by further lowering the external magnetic field $B$. We ascribe this unexpected cross-over to vortex channeling due to a fan-like penetration of the applied magnetic field as confirmed by magneto-optic imaging. The observation of $B_{cr1}$ becomes a direct evidence of a general feature in type-II superconducting films at low fields, that is a channel-like vortex motion induced by the inhomogeneous magnetic state caused by the relatively strong pinning.
\end{abstract}

\pacs{74.25.F-, 74.25.Wx, 74.78.-w}

\keywords{vortex dynamics; flux flow instability; Type-II superconducting thin films}

\maketitle
	
%Body of paper
\section{\label{sec:Intro}Introduction}
In type-II superconductors non linear effects of the flux flow dynamics of Abrikosov vortices driven by high bias currents have been deeply studied in the past in the particular case of homogeneous state in weak pinning materials, regardless the pinning mechanism of the vortex-defect interaction.\cite{LO-Book1986, LO-JLTP1979, Klein-JLTP1995} On the contrary, an attractive interest has been recently devoted to different superconducting materials in which inhomogeneous states are induced by the type of pinning, intrinsic or extrinsic, causing nonlinearities of the flux flow motion resulting in a preferential as well as a directional motion of vortices.\cite{Karapetrov-PRB2009, Silhanek-SST2009, Verellen-APL2008, Kokubo-PRB2006, Villegas-PRB2005}\\
At high driving currents the current-voltage (I-V) characteristics show an abrupt switching from the linear flux flow state into the normal state at a threshold voltage $V^{*}$. This voltage jump can be accounted for different intrinsic or extrinsic mechanisms such as for example Joule heating.\cite{Samoilov-PRL1995, Xiao-PRB1998} An intrinsic non linear conductivity based on non equilibrium effects has been predicted by Larkin and Ovchinnikov (LO).\cite{LO-Book1986, LO-JLTP1979} In this approach a sufficiently high vortex velocity $v$ determines the escape of quasiparticles from the vortex core. The finite energy relaxation time $\tau_{e}$ of quasiparticles leads to appreciable changes of the quasiparticles distribution and to the shrinking of the vortex core.\cite{LO-Book1986, LO-JLTP1979} The vortex velocity dependence of the flux flow resistivity implies an instability of the moving vortex lattice at a threshold velocity $v^{*}$ with a sudden jump to the normal state.\\
Within the LO approach, in a magnetic field range in which a spatially uniform quasiparticle distribution can be assumed \cite{LO-Book1986}, a magnetic field independent critical velocity $v^{*}_{\textrm{LO}}$ is derived as function of the reduced temperature $t = T / T_{c}$:\cite{LO-Book1986}
	\begin{equation} \label{eq:vLO}
	v^{*}_{\textrm{LO}}(t) = \frac{D^{1/2} \left[ 14 \zeta(3) \right]^{1/4} \left( 1 - t \right)^{1/4}}{\left( \pi \tau_{e} \right)^{1/2}} \, .
	\end{equation}
\noindent Here $D$ is the quasiparticle diffusion coefficient and $\zeta(x)$ is the Riemann zeta function. The LO prediction has been confirmed on several superconducting materials and multilayers above a cross-over field $B_{cr}$, while below $B_{cr}$, the power-law dependence $v^{*} \approx B^{-1/2}$ was found.\cite{Peroz-PRB2005, Angrisani-PRB2007, Doettinger-PhC1995, Doettinger-PRB1997, Kalisky-PRL2006} This well known cross-over effect has been justified to ensure the spatial homogeneity of the non-equilibrium  quasi-particles distribution in low magnetic fields.\cite{Doettinger-PhC1995}\\
In this paper we focus on a very low field region in wide Nb strips, in which we identify a new cross-over in the magnetic field dependence of the vortex critical velocity $v^{*}$. We study this field regime by both I-V and Magneto-Optical-Imaging (MOI) measurements in order to elucidate the role of the first flux penetration in the high velocity vortex dynamics.\\
The observation of a new cross-over field, $B_{cr1}$, from the regime where $v^{*}$ increases with decreasing magnetic field ($v^{*} \approx B^{-1/2}$) to another regime where $v^{*}$ decreases with further decreasing the field, can be ascribed to an inhomogenous state induced by the flux penetration. In fact, in the very low field region, $B \leq B_{cr1}$, the flux does not penetrate with a smooth advancing front, but instead as a series of irregularly shaped protrusions,\cite{James-PhC2000} causing the formation of an inhomogeneous magnetic field distribution as observed by MOI.\cite{Altshuler-RMP2004,*Jooss-RPP2002, Gozzelino-SST2003,*Laviano-SST2003} This kind of flux penetration is compatible with the presence of circular defects or micro-pores, as revealed by a FESEM analysis, which shows they are large enough to preferentially channel the moving vortices.

\section{\label{sec:Exp}Experiment}
\begin{figure}[!t]
	\begin{center}
	\includegraphics[width=0.45\textwidth]{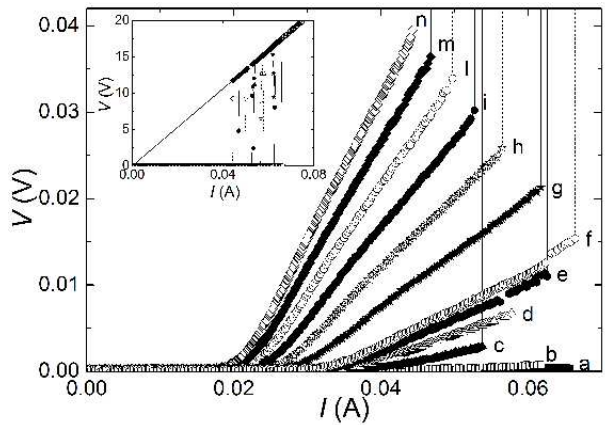}
	\caption{\label{fig:1}I-V curves for different values of $B$ in the low field range, at $T = 4.2$~K, for NbA1 sample. The field values are: a=0, b=0.78, c=1.56, d=2.08, e=2.34, f=2.60, g=3.90, h=5.20, i=6.50, l=7.80, m=9.10, n=10.40~mT. Inset: Full view of the I-V curves where the transition to the normal state is shown.}
	\end{center}
	\end{figure}

\begin{figure}[!t]
	\begin{center}
	\includegraphics[width=0.45\textwidth]{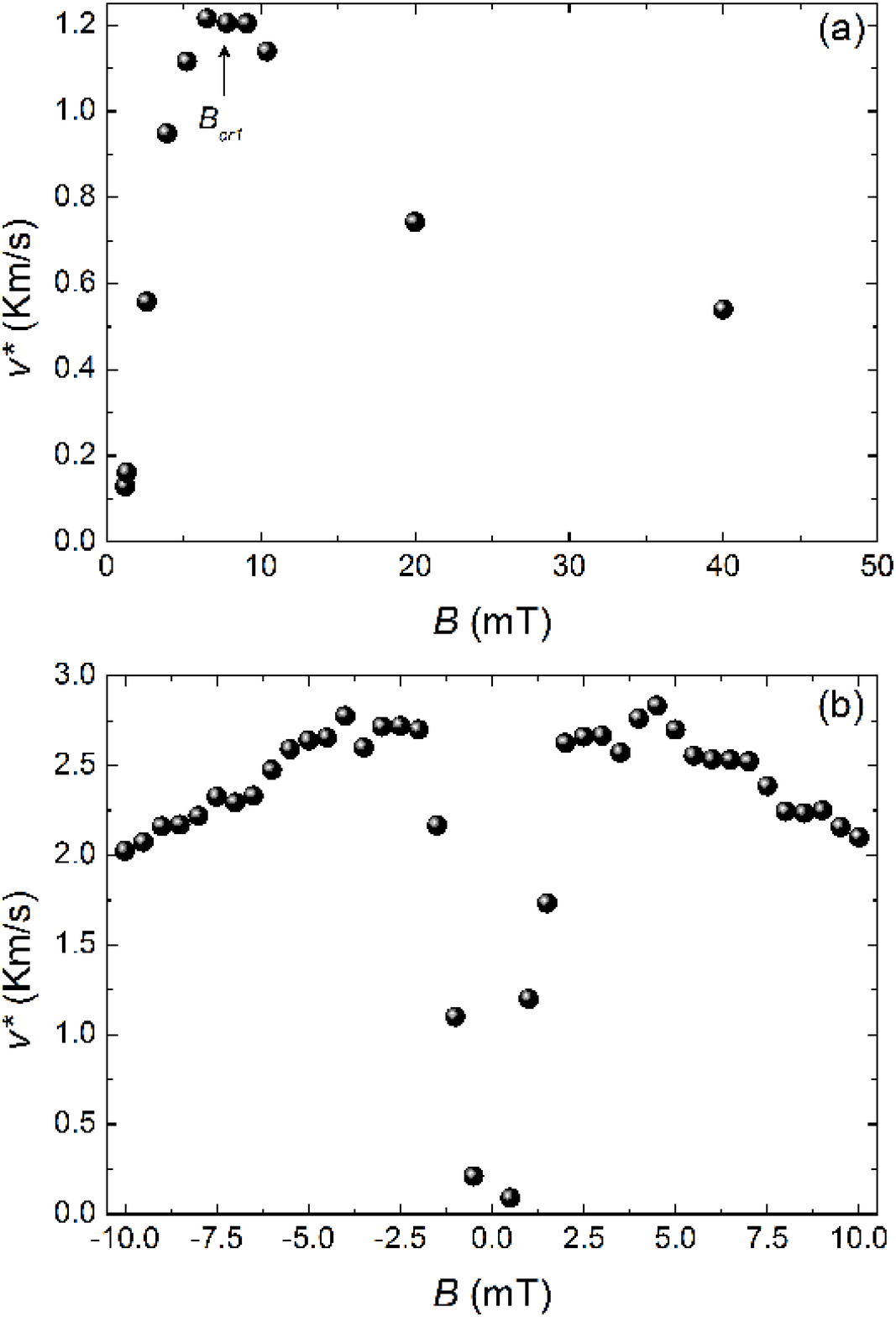}
	\caption{\label{fig:2}a) Vortex critical velocity $v^{*}$ as function of $B$ in the low-field range, at $T = 4.2$~K for the sample NbB5. The arrow marks the crossover field $B_{cr1}$. b) The $v^{*}(B)$ dependence at $T = 4.2$~K for the sample NbA02 in the low-field range for positive and negative field values.}
	\end{center}
	\end{figure}

\begin{figure}[!t]
	\begin{center}
	\includegraphics[width=0.45\textwidth]{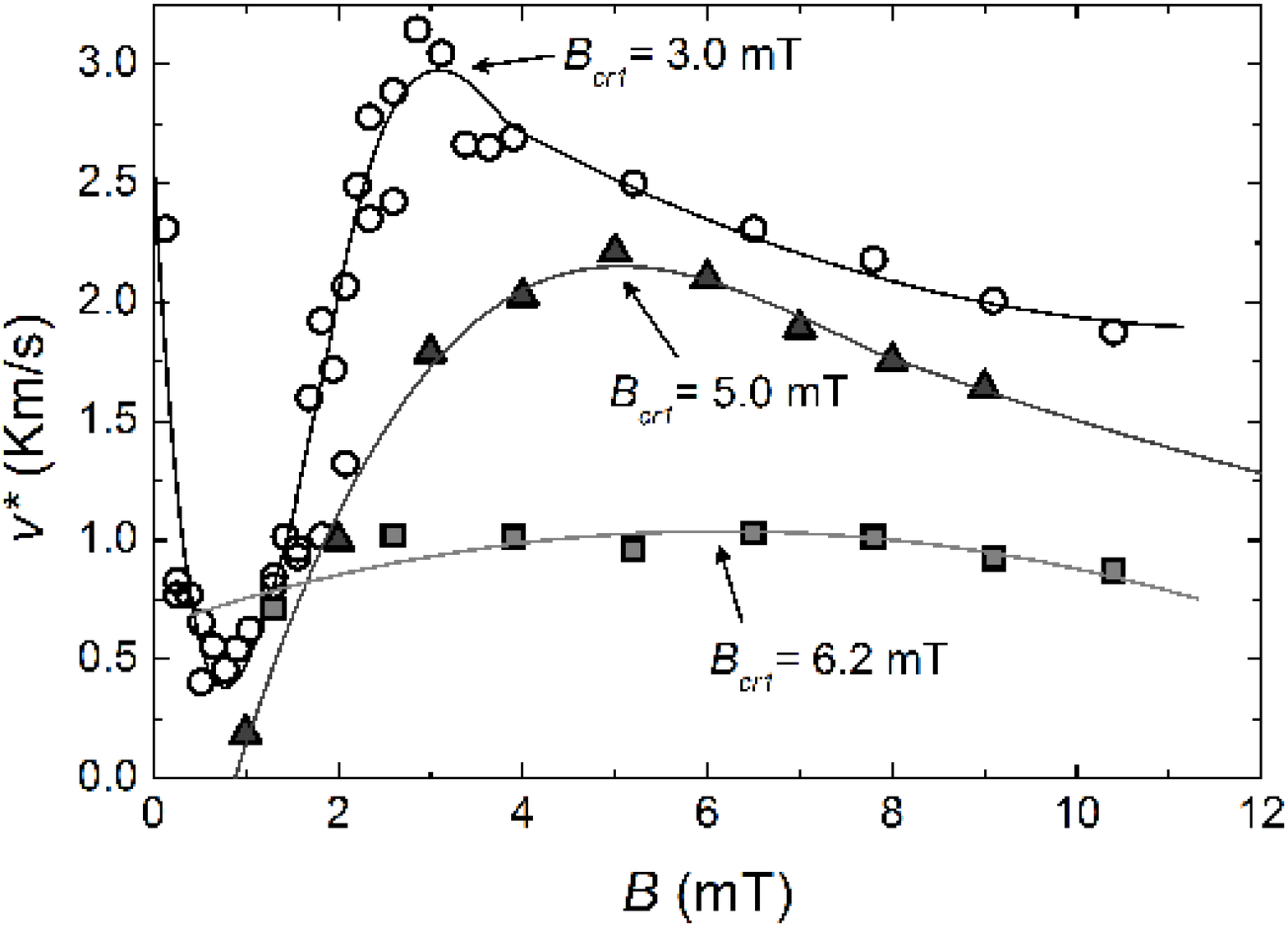}
	\caption{\label{fig:3}Critical vortex velocity as function of $B$ in the low field region, at $T = 4.2$~K, for three different strips on the 60~nm thick Nb film. Open circles refer to NbA1 ($w = 20~\mu$m),triangles to NbA3 ($w = 50~\mu$m), and squares to NbA4 ($w = 100~\mu$m). The arrows mark $B_{cr1}$ for each sample.}
	\end{center}
	\end{figure}

\begin{figure}[!t]
	\begin{center}
	\includegraphics[width=0.45\textwidth]{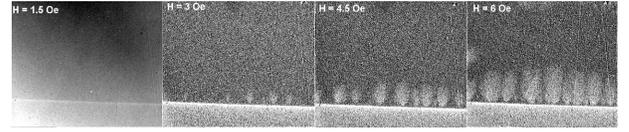}
	\caption{\label{fig:5}MOI measurements of the sample NbA4 at $T = 4.1$~K. After zero field cooling, a sequence of increasing magnetic field was applied. Dark contrast represents the zero-field Meissner state. The bright line is the sample edge along the strip length. A fan-like penetration pattern clearly develops after vortex nucleation occurred at sample edges.}
	\end{center}
	\end{figure}

We performed I-V measurements by the standard four contact technique on Nb films of different thickness in the range $d = 30 \div 150$~nm. The samples were grown on Si(100) substrates by a UHV dc diode magnetron sputtering. The films were deposited at typical rates of 0.28~nm/s. The Nb strips were obtained by standard photolithographic technique which provides four samples of different widths ($w = $20, 40, 50, 100~$\mu$m) on the same substrate, with a length between the voltage contacts $L = 2$~mm. All samples are characterized by an RRR($ = \rho_{300\textrm{K}}/\rho_{10\textrm{K}}$) $\geq 2$. The Nb films are characterized by a Ginzburg-Landau parameter $\kappa \geq 6$, being the coherence length $\xi_{0} = 7$~nm and the electron mean free path $l = 5$~nm with the typical value of the penetration depth $\lambda_{L} = 39$~nm,\cite{Grimaldi-PRB2009} so they show type-II superconducting properties. The sample parameters are summarized in the Table I.
	\begin{table}[h]
	\caption{\label{tab:samples}Physical parameters of the samples}
		\begin{ruledtabular}
			\begin{tabular}{lccccc}
			Sample & $d$ (nm) & $w$ ($\mu$m) & $T_{c}$ (K) & $\Delta T_{c}$ (K) & $\rho_{10\textrm{K}}$ ($\mu \Omega \cdot$cm)\\
			\hline
			NbA02 & 30 & 40 & 7.6 & 0.2 & 24.0\\
			NbA04 & 30 & 100 & 7.6 & 0.2 & 25.0\\
			NbA1 & 60 & 20 & 8.2 & 0.3 & 15.5\\
			NbA3 & 60 & 50 & 8.1 & 0.2 & 16.5\\
			NbA4 & 60 & 100 & 8.2 & 0.2 & 16.5\\
			NbB5 & 100 & 100 & 8.5 & 0.1 & 15.3\\
			NbC6 & 135 & 50 & 8.7 & 0.1 & 16.5\\
			NbD8 & 150 & 50 & 9.2 & 0.2 & 9.5\\
			\end{tabular}
		\end{ruledtabular}
	\end{table}\\
\noindent We investigated a temperature range down to 0.5~$T_{c}$ and a broad magnetic field interval, perpendicularly applied to the film surface. The superconducting transition temperature $T_{c}$ is the value corresponding to $R \leq 10^{-4}~\Omega$, well below the thresholds used to compute the transition width $\Delta T_{c}$, estimated by the criterion from 10\% to 90\%. The slightly high values of $\rho_{10\textrm{K}}$ are related to the fabrication process. The typical value of the critical current density at $T = 4.2$~K in zero external field is $J_{c} \approx 3 \cdot 10^{10} \textrm{A}/\textrm{m}^{2}$. In our experiments we have previously performed a systematic study on the influence of self-heating by changing the bias current operation mode in order to ensure that the observed flux flow instability (FFI) is not triggered by Joule self-heating, as reported elsewhere.\cite{Grimaldi-PRB2009, Grimaldi-JPCS2008} In order to minimize the unavoidable heating effects at high driving currents, rectangular pulses have been used with a current-on equal to 2.5 ms and a current-off equal to 3.5 ms. We also compare our experimental I-V data with the numerical results obtained by J. Maza et al.\cite{Maza-PRB2008} Main differences are that the calculations have been performed for $T$ close to $T_{c}$ with no applied magnetic field, and the numerical building procedure of the I-V curve is made up by a staircase ramp of current, which is not the bias operation mode we have experimentally performed. Instead, to minimize self-heating, we have chosen the short pulses mode, each pulse of length equal PW with a delay equal to PD, as described elsewhere.\cite{Grimaldi-JPCS2008} We have studied the influence of the pulse width and pulse delay on the I-V measurements,\cite{Grimaldi-JPCS2008} thus we established that PW = 2.5 ms is the maximum value and PD = 3.5 ms is the minimum value which are compatible with a reproducible I-V curve, so that the same instability point $(I^{*}, V^{*})$ is reached. By sweeping upward and downward the pulse amplitude, no hysteresis in I-V curves has been observed. From standard thermal analysis and from the magnetic field dependent dissipation power $P^{*} = I^{*}\,V^{*}$ ($I^{*}$ is the current value at the instability point), we excluded that the observed threshold voltages $V^{*}$ were determined by heating mechanisms.\cite{Grimaldi-JPCS2008} Indeed, if it is the case of thermal runway, such $P^{*}$ should result magnetic field independent.\cite{Xiao-PRB1999}\\
We also underline that I-V measurements have been performed following both Zero Field Cooling (ZFC) and Field Cooling (FC) procedures, leading to the same results, as expected, since for each I-V curve, data have been recorded by driving the system up to the normal state with sufficiently high bias current.\\
I-V curves of the sample NbA1, for different values of the magnetic field at $T = 4.2$~K, displaying the voltage jumps, are shown in Fig.~\ref{fig:1}. In particular the I-V curves for fields below $B \leq 10.0$~mT have been investigated with a fine variation of the external field around $0.1$~mT. From the measured I-V data, the critical vortex velocity $v^{*}(B) = V^{*} / B\,L$ has been deduced for each magnetic field value.\cite{Tinkham-Book1996}
\noindent The magnetic field dependence of $v^{*}$ is presented in Fig.~\ref{fig:2}a and \ref{fig:2}b, for the NbB5 and NbA02 samples respectively, in the low field range $B < 50.0$~mT. Whereas for larger fields, it has been clearly observed the usual cross-over field $B_{cr}$ from the regime where $v^{*}$ is proportional to $B^{-1/2}$, to another regime at higher fields where $v^{*}$ is found almost constant.\cite{Grimaldi-PRB2009, Grimaldi-PhC2008}. From Fig.~\ref{fig:2}a, surprising in the lowest field regime, for the NbB5 sample, it is evident the presence of a new cross-over at a magnetic field value $B_{cr1} \approx 9$~mT, below which $v^{*}$ decreases with decreasing the external field. In Fig.~\ref{fig:2}b, for the NbA02 sample, the $v^{*}(B)$ dependence is displayed in the low field range for both positive and negative field values. In this case it is clear that the critical vortex velocity goes to zero crossing the zero magnetic field value, as well as the $v^{*}(B)$ curve is symmetric with respect to zero field.
\noindent We outline that all the measured samples show the same remarkable feature: the presence of a new cross-over field $B_{cr1}$ in the low field region, regardless the thickness of the Nb film and the strip width. However, a study of this unusual low field behavior on the same Nb film for different line widths of the strip is presented in Fig.~\ref{fig:3}.
	\begin{figure}[!b]
	\begin{center}
	\includegraphics[width=0.45\textwidth]{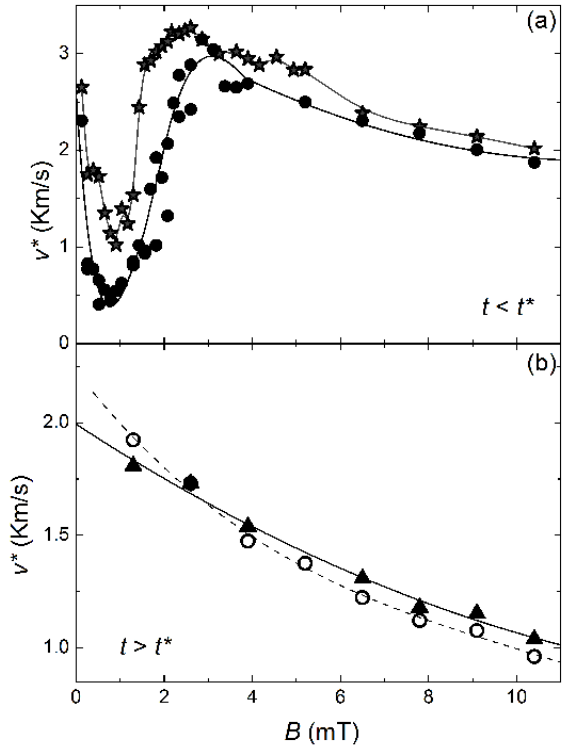}
	\caption{\label{fig:4}Critical vortex velocity as function of $B$ in the low field region for NbA1, at different $t = T/T_{c}$ a) At $t =$~0.5~(full circles) and 0.6~(stars). b) At $t =$~0.90 (open circles) and $t = $~0.99 (full triangles).}
	\end{center}
	\end{figure}
\noindent It is shown that the $B_{cr1}$ cross-over value is shifted to lower fields as the line width becomes smaller. This means that to be able to detect such experimental behavior, not only a fine variation of the external magnetic field should be ensured, but also the line width should be sufficiently large. Moreover, in principle, no upper limit to the line width can be established. This clearly states a size dependence of the cross-over low field.\\
The temperature behavior of the $B_{cr1}$ cross-over field has also been investigated. In Fig.~\ref{fig:4}a the evolution of the $v^{*}(B)$ for different temperatures up to $T_{c}$ is reported for the NbA1 sample. A threshold temperature $t^{*} = 0.75$ is found, below which the anomalous low field behavior of $v^{*}(B)$ is preserved, with a small variation of the $B_{cr1}$ value, between 2.5 and 3~mT. On the contrary, as shown in Fig.~\ref{fig:4}b, above $t^{*}$ the $B_{cr1}$ crossover field disappears, leading to a monotonous decreasing function of $v^{*}(B)$. Similar results have been obtained on other samples.\cite{Grimaldi-JPCM2009}\\
In order to obtain more information on the observed anomalous low field behavior $v^{*}(B)$, we visualized the first flux penetration by MOI technique with an indicator film.\cite{Altshuler-RMP2004} A description of the experimental set-up can be found elsewhere.\cite{Gozzelino-SST2003} The results of the measurements can be here summarized:\\
1) the first penetration of vortices occurs in a discontinuous way, preferably through small circular defects along the sample edges;\\
2) no dendritic pattern is observed with both transport current up to 40~mA and applied magnetic field up to 80~mT;\\
3) the flux penetration is characterized by a typical fan-like shaped pattern, as shown in Fig.~\ref{fig:5}, which becomes smoother and smoother as the vortices diffuse deeper in the film.
	\\
\noindent In addition we have studied the morphology of the Nb strips by a FESEM analysis, which confirms the presence of circular defects both along the edges and the inner part of the strips, randomly distributed, whose size results in few tenth of micron width. We estimate that micro-pores width is around $0.4~\mu$m, and their spacing is almost $5~\mu$m, within a cluster of around $20~\mu$m, then clusters are randomly distributed in each strip.

\section{\label{sec:Disc}Discussion}
The low field dependence of $v^{*}(B)$ can be understood taking into account the non uniform magnetic flux penetration, with the protrusion formation as the field starts to penetrate in the mT range (see Fig.~\ref{fig:5}). In fact the fan-like magnetic flux penetration mirrors the non homogeneous state of the fairly strong pinning Nb strips, as already outlined in the literature. \cite{James-PhC2000,Gozzelino-SST2003}\\
Moreover, by applying the driving current, in the field penetrated regions a flux flow state of the moving vortices is achieved. MOI and direct transport measurements give us complementary information: the former states the formation of channels due to the fan-like penetration, the latter states that vortices initially penetrated in the channels, can be driven afterwards at high velocities. Therefore the inhomogeneous magnetic penetration induces preferential stationary channels for the moving vortices with a total effective width $l$. In particular, from FESEM analysis, these channels can be assumed around $0.4~\mu$m wide, comparable to the micro-pores, grouped together in clusters along the strip. This means that the voltages measured across the distance $L$ between the voltage tips are actually produced across an effective distance $l < L$; so that in the low field range, reasonably for $B < B_{cr1}$, the flux flow channels fill only partially the whole strip length.\\
For this reason an analogous scale factor should affect all the voltage measurements and in particular the vortex critical velocity and the flux flow resistance.
	\begin{figure}[!t]
	\begin{center}
	\includegraphics[width=0.45\textwidth]{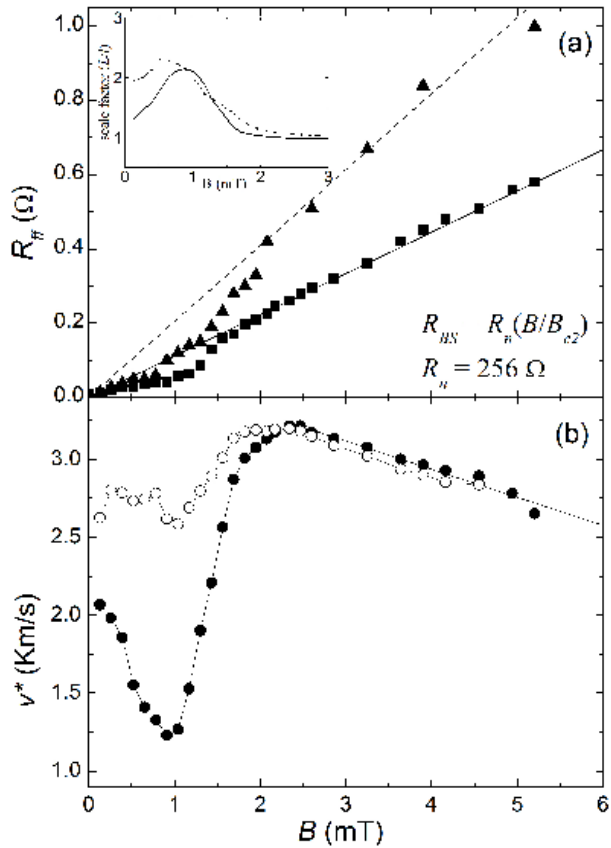}
	\caption{\label{fig:6}a) Flux flow resistance as function of $B$ in the mT range for NbA1, at $t = 0.67$ (squares) and $t = 0.74$ (triangles). The corresponding lines refer to the Bardeen-Stephen prediction $R_{BS} = R_{n} B / B_{c2}$. Inset: Scale factor $L/l$ as function of $B$ at $t = 0.67$ (solid line) and $t = 0.74$ (dotted line). b) Vortex critical velocity as function of $B$ at $t = 0.67$ for NbA1 corrected by the factor $L/l$ (open circles). The full circles represent the experimental data.}
	\end{center}
	\end{figure}
\noindent On the contrary when the flux penetration becomes smooth and regular Bean-like, above the threshold temperature $t^{*}$, the pinning effectiveness decreases so the vortex flow becomes uniform as well. Therefore at temperatures below $t^{*}$ and in the mT range, the flux flow resistance $R_{ff}$ is expected to assume lower values than the Bardeen-Stephen prediction (RBS), indeed this is shown in Fig.~\ref{fig:6}a. Our hypothesis is sustained by the observed linear dependence of $R_{ff}$ above the same threshold temperature $t^{*} \approx 0.75 T_{c}$. Linear fits of the I-V curves close to $V^{*}$ have been performed in order to get the dynamic resistance in the flux flow state, $R_{ff}$, which are displayed in Fig.~\ref{fig:6}a. Inset of Fig.~\ref{fig:6}a reports the scale factor $L/l$ as function of $B$, as deduced by our data. The scale factor displays a maximum in the same low field range in which the flux flow resistance $R_{ff}$ shows a minimum (see Fig.~\ref{fig:6}a). Since $R_{ff}$ turns equal to $R_{BS} = R_{n} B / B_{c2}$ ($B_{c2}$ has been determined with the criterion of $J_{c}(B_{c2}) = 0$ from the corresponding $J_{c}(B)$), $L/l$ becomes equal to 1 around $B \approx 2$~mT, which is the same cross-over field $B_{cr1}$ found in the corresponding $v^{*}(B)$ curves of Fig.~\ref{fig:4}a.\\
In Fig.~\ref{fig:6}b we report at $t = 0.67$ the magnetic field dependence of $v^{*}$ corrected by the $L/l$ factor for the NbA1 sample. At low fields, $v^{*}(B)$ is still well different from the usually observed power-law dependence $v^{*} \approx B^{-1/2}$. Nevertheless the very low field dependence of $v^{*}$ has not yet been studied in the literature so far. In this field range, in the case of isolated vortices, a value of the critical velocity for $B$ towards zero, $v^{*}_{0}$, larger than the LO prediction should be found. As the field increases from zero up to values such that the vortex lattice constant $a_{0}=(\frac{2}{\sqrt{3}} \frac{\Phi_{0}}{B})^{1/2}$ is larger than the quasi-particle diffusion length $l_{e} = (D \tau_{e})^{1/2}$, the damping coefficient $\eta$ of the moving vortex lattice is equal to the one for isolated vortices. For this reason $v^{*}$ should approach an almost constant value $v^{*}_{0}$. Only for larger fields, i.e. for $a_{0} < l_{e}$ , as for as $B < B_{cr}$ the power-law $B^{-1/2}$ dependence should be recovered; finally for $B > B_{cr}$ the LO prediction is followed.\cite{Grimaldi-PRB2009} In our case we evaluate the quasiparticle diffusion coefficient $D = (\frac{1}{3} v_{F} l_{e}) = 2.0 \times 10^{-4} \textrm{m}^{2}/\textrm{s}$ (using for the Nb Fermi velocity the value $v_{F} = 2.73 \times 10^{5}$~m/s [\onlinecite{Kerchner-PRB1981}] and deducing the electronic mean free path, $l_{e}$, from the measured low temperature resistivity [\onlinecite{Hauser-PR1964}]) and $\tau_{e} = 5 \times 10^{-10}$~s fitting eq. (1),\cite{Grimaldi-PRB2009} so that the condition $a_{0} \approx l_{e}$ corresponds to $B^{*} \approx 39$~mT. This means that in the low-field range considered in this work neither the LO nor the $B^{-1/2}$ predictions should be expected. Nevertheless, to give a quantitative estimate of the low field cut-off vortex critical velocity $v^{*}_{0}$, the contribution from the fan-like penetration should also be added. Indeed from MOI measurements it is evident that vortices entering the sample have a non-vanishing velocity component perpendicular to the Lorentz force (i.e. along the direction of the applied bias current), thus leading to lower local values of the longitudinal component of the flux flow critical velocity $v^{*}$. Therefore, the almost constant value found experimentally for $B$ towards zero in Fig.~\ref{fig:6}b, can be further corrected to the actual value of $v^{*}_{0}$. Moreover, for a quantitative analysis the non uniform self field generated by the bias current and local inhomogeneities within the channels should also be considered. On one hand, a dramatic influence of self-field effects, which could be comparable with the lower applied fields, can be excluded. In fact for our samples at low magnetic fields the depinning critical current $I_{c}$ has a steep decrease as function of the external $B$ which therefore is dominant with respect to the self field. On the other hand, a more realistic non constant width of the vortex channels determines a non-homogeneous distribution of vortex velocity. As a consequence locally the critical vortex velocity can be reached before the mean critical vortex velocity, so this implies a further reduction of the measured threshold voltage $V^{*}$.\\
In the end, a possible contribution of vortex-antivortex (V-AV) pairs to the vortex critical velocity at very low fields should be also considered. In particular such an effect becomes dominant at $B = 0$ and close to $T_{c}$ in the presence of a current biasing. In principle V-AV pairs formation exist, but to be significant a saturation of the critical voltage $V^{*}$ to a finite value at low fields should occur. By comparing our data on several Nb samples, we can reasonably state that no V-AV pairs contribution to $v^{*}(B)$ is usually observed, see for example Fig.s~\ref{fig:2}a~and~\ref{fig:2}b, where the critical velocity $v^{*}$ approaches zero value crossing the $B = 0$ axis. Nevertheless in Fig.s~\ref{fig:4}~and~\ref{fig:6}b an almost constant value $v^{*}_{0}$ is found for $B \sim 0$, so that V-AV pairs may contribute at least in the case of the smallest line width investigated.\\
Finally, we emphasize that, very recently, the observation of this new cross-over effect has been confirmed in Al thin films grown on top of a periodic array of magnetic micro-loops,\cite{Leo-PhCVortex} if magnetized in a flux-closure state.\cite{Silhanek-PhC2008} In this case the magnetic state of the artificial pinning structure is also effective to sustain a channel-like vortex motion.\cite{Verellen-APL2008} A comparison with the rather strong intrinsic pinning Nb superconductor outlines that pinning, intrinsic or extrinsic, significantly affect the vortex critical velocity in type-II superconducting thin films.\\
In summary a major cause of the observed reduced flux flow resistance at low applied magnetic fields is very likely the inhomogeneous flux penetration due to the micro-porosity, which is mirrored in the unusual low field behaviour of the vortex critical velocity $v^{*}(B)$.

\section{\label{sec:Conclusions}Conclusions}
We observe a low field cross-over effect in the magnetic field dependence of the critical vortex velocity in wide Nb strips. We believe that this cross-over effect in the low-field regime can be justified by vortex channeling through the inhomogeneous pinning center distribution, as emphasized by MOI measurements. We observe that $B_{cr1}$ increases by both increasing the line width and the thickness of the samples. This suggests that such cross-over effect can be hidden by the sample size.\\
The channeling effect on the moving vortices is further confirmed by the magnetic field dependence of the flux flow resistance which turns back into the Bardeen-Stephen linear behavior as the $B_{cr1}$ field is overcame. Therefore $B_{cr1}$ marks the crossover between the channel-like vortex motion and the uniform flux flow state. This observation is likely to be a common feature in type-II superconducting thin films exhibiting moderately strong pinning, as preliminary observed in the case of Al thin film grown on artificial pinning structures,\cite{Leo-PhCVortex} and here clearly established for the intrinsic Nb superconductor.

%Acknowledgments
\begin{acknowledgments}
We are grateful to C. Attanasio for useful discussions, C. Cirillo for Nb thin films deposition, and to G. Carapella for providing FESEM images. This work was partially supported by the Research Project L.R. N°5, Regione Campania. A. L. acknowledges the funding support under the contract CO 03/2009.
\end{acknowledgments}

%Bibliography
%\bibliography{Grimaldi_et_al-date_31-03-2010}

\end{document}